# Skyrmions as Compact, Robust and Energy-Efficient Interconnects for Domain Wall (DW)-based Systems

Wang Kang, *Member, IEEE,* Yangqi Huang, Xichao Zhang, Yan Zhou, Weifeng Lv and
Weisheng Zhao*, *Senior Member, IEEE*

*Abstract*—Magnetic domain-wall (DW) has been widely investigated for future memory and computing systems. However, energy efficiency and stability become two major challenges of DW-based systems. In this letter, we first propose exploiting skyrmions as on-chip and inter-chip interconnects for DW-based systems, owing to the topological stability, small size and ultra-low depinning current density. In the proposed technique, data are processed in the form of DWs but are transmitted instead in the form of skyrmions. The reversible conversion between a skyrmion and a DW pair can be physically achieved by connecting a wide and a narrow magnetic nanowire. Our proposed technique can realize highly compact, robust and energy-efficient on-chip and inter-chip interconnects for DW-based systems, enabling the system to take advantages of both the DW and skyrmion.

## I. INTRODUCTION

SPINTRONIC domain wall (DW) has been widely considered as a promising candidate in future memory and computing systems [1-3]. Until now, huge academic and industrial efforts have been captured in this area and quite a few fruits have been achieved, e.g., racetrack memory [4] and logic design [5]. The fundamental advantages of DW are its intrinsic non-volatility, high density, good retention and endurance etc. However, the DW motion (i.e., data transmission) requires a relatively large critical current density ($10^{10} - 10^{12}$ A/m$^2$), leading to energy inefficiency, especially for long distance on-chip or inter-chip interconnects [6]. In addition, DW motion faces stability issues in the presence of process defects or imperfections [7]. These challenges significantly limit the applications of utilizing DW for long distance data transmission (see Fig. 1(a)). One possible strategy (see Fig. 1(b)) is to firstly transform DW into electrical current (or voltage) at the transmitter side, and then transmit the electrical signals through a metal channel (e.g., Cu), finally transform the electrical signal back to the form of DW at the receiver side for further data processing. This strategy involves writing and reading DW as well as peripheral control circuits, resulting in even increases of energy and design complexity. In addition, this strategy undermines the non-volatility features of the DW-based system, as data may lost during the transmission process if the supply power is accidently off.

Recently, skyrmions, particle-like topologically stable spin configurations, have attracted a lot of attention [8-9]. They hold great promise as information carriers in the next-generation spintronic devices, owing to the topological stability, small size and extremely low driving current density needed to move them. Compared with DW, DW is suitable for data processing as it facilitates conversion with electrical signals, whereas skyrmion is superior for data transmission, because: (a) the skyrmion size and spacing (down to a few nanometers) are much smaller than those (hardly below 30-40 nm) of DW, allowing higher data throughput; (b) the critical depinning current density ($10^6 - 10^{10}$ A/m$^2$) of skyrmion is 2-4 orders lower than that of DW, resulting in much more energy-efficient data transmission for low-power applications; (c) skyrmion is protected topologically against dissipation and fluctuation in the presence of process defects or imperfections during the motion, enabling more robust data transmission [9-11]. Therefore, skyrmion processes the potential to outperform DW for data transmission with higher stability, throughput and energy-efficiency.

In this letter, we explore the favorable features of skyrmion for on-chip and inter-chip interconnects for DW-based systems. We show the reversible conversion between a skyrmion and a DW by varying the width of the magnetic nanowire, achieving highly compact realization. Furthermore, design considerations and performance evaluation are performed. With the proposed technique, data are processed in the form of DW but transmitted in the form of skyrmion, enabling the system to take advantages of both the DW and skyrmion.

## II. SKYRMIONS AS INTERCONNECTS FOR DW SYSTEMS

A skyrmion is characterized by the skyrmion number ($Q$), which is the topological number in the planar system. In general, we have $|Q| = 1$ for a skyrmion in a sufficiently large area and it keeps a constant as the skyrmion continuously deforms (topologically protected). On the other hand, we have $|Q| = 0$ for the DW [11]. In most systems, the spin configurations of a

Manuscript received January 10, 2016. This work was supported in part by the China Postdoctoral Science Foundation (2015M570024), and the National Natural Science Foundation of China (61501013 and 61571023).

Wang Kang an Weifeng Lv are with the Spintronics Interdisciplinary Center and School of Computer Science and Engineering, Beihang Univeristy, Beijing, 100191, China (e-mail: wang.kang@buaa.edu.cn).

Xichao Zhang and Yan Zhou are with the Department of Physics, University of Hong Kong, Hong Kong, China (e-mail: yanzhou@hku.hk).

Yangqi Huang and Weisheng Zhao are with Spintronics Interdisciplinary Center, Beihang Univeristy, Beijing, 100191, China (corresponding author: Weisheng Zhao; phone: +86-010-82314875; fax: +86-010-82339374; e-mail: weisheng.zhao@buaa.edu.cn).

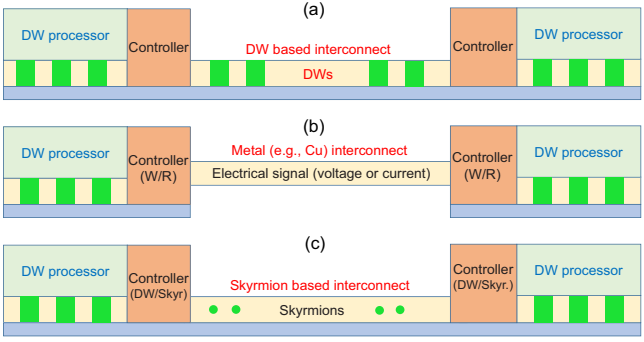

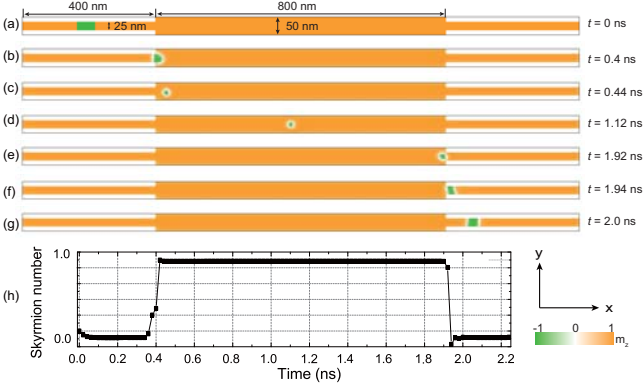

Fig. 1. Illustrations of the (a) DW-based interconnect; (b) metal interconnect; and (c) proposed skyrmion-based interconnect for DW-based systems.

Fig. 2. Illustration of reversible conversion between a skyrmion and a DW pair. (a)-(g) The snapshots of magnetization configuration at various selected times; (h) The time evolution of the skyrmion number ($Q$).

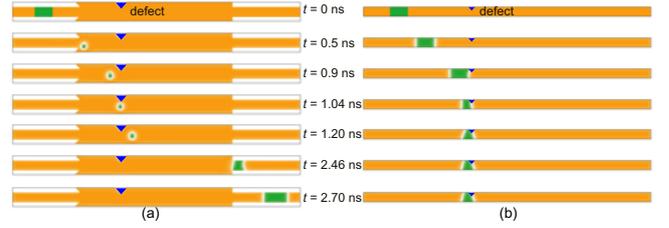

Fig. 3. Illustration of the impact of pinning defects (triangular red region) on the (a) skymion and (b) DW motion, respectively.

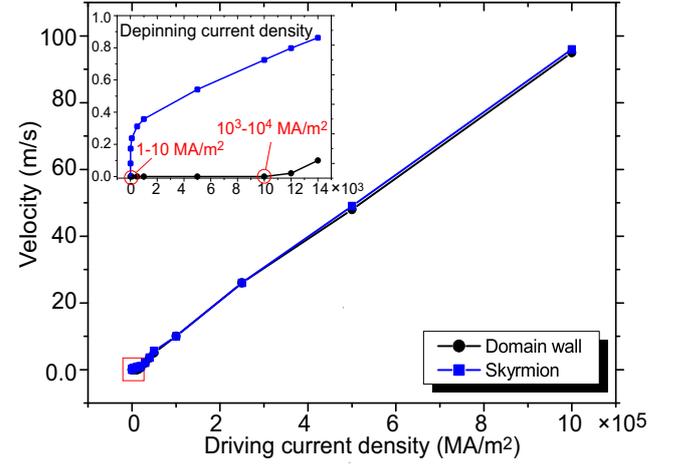

Fig. 4. The DW and skyrmion motion velocities with respect to the driving current density; (inset) The depinning current densities of skyrmion and DW.

skyrmion originate from the chiral exchange interactions, i.e., Dzyaloshinskii-Moriya interaction (DMI), in lattices or at the interface of magnetic films lacking inversion symmetry [12]. For magnetic thin films (main focus of this letter), the material system of skyrmion is very similar to that of DW, e.g., CoFeB or Co/Pt [10]. Recently, chiral DW with a preferred handedness (influenced by DMI) have also been observed [13, 14]. As we know that one critical challenges of DW-based systems is to further reduce the energy consumption for DW motion whereas skyrmion holds great promise as information carriers for low power data transmission. Therefore, this challenge can be well solved if we can integrate DW and skyrmion in one system, which explores advantages of both the DW and skyrmion.

Fig. 1(c) illustrates the schematic of utilizing skyrmions as on-chip and inter-chip interconnects for DW-based systems, in which data are processed in the form of DWs but transmitted in the form of skyrmions. First of all, we investigate the mutual conversion between a skrymion and a DW pair in a magnetic nanowire (e.g., CoFeB or Co/Pt with perpendicular magnetic anisotropy, PMA) with variable widths (see in the next Section), as shown in Fig. 2(a). Generally, the intrinsic diameter ($R$, here $R \approx 25\ nm$ with the default parameters) of a skyrmion is mainly determined by the material parameters and the nanowire geometry [11]. In our configuration, the nanowire width is set to be $W \leq R$ (e.g., 25 nm) at the two ends but with $W \geq 2R$ (e.g., 50 nm) at the center. In this case, a skyrmion can move stably along the nanowire for $W \geq 2R$, alternatively it is transformed to be a DW for $W \leq 2R$. Fig. 2(a)-(h) illustrates the dynamic evolutions between a skyrmion and a DW. At the initial state, we nuclear a DW at the left side of the nanowire (see Fig. 2 (a)). By applying a spin-polarized driving current (through either spin transfer torque, STT, or spin Hall effect, SHE), the DW moves along the nanowire from the left to the right side. When the DW arrives at the left junction (at time $t = 0.4\ ns$) of the nanowire, it deforms into a skyrmion after entering the wide part of the nanowire (see Fig. 2(b)-(c)). This process is very like blowing bubbles with a straw. To accelerate the transformation process, we further designed a 45 degrees interface angle at the junction (see Fig. 3(a)). After that, the skyrmion moves stably along the nanowire (see Fig. 2(d)) until it reaches the right junction of the nanowire (at $t = 1.92\ ns$, see Fig. 2(e)). As the width of the narrow part of the nanowire is smaller than $2R$, the skyrmion cannot stably exist in this case. As a result, the skyrmion turns into a DW and continuously moves on (see Fig. 2(f)-(g)). In addition, Fig. 2(h) shows the time evolution of the skyrmion number ($Q$). We can observe that the skyrmion number changes from 0 to 1 as the DW deforms into a skyrmion; otherwise it changes from 1 to 0. As can be seen, the revisable conversion between a skyrmion and a DW pair can be physically achieved by varying the geometry of the nanowire, making the system be highly compact.

As skyrmion is topologically protected once it is created, it is more robust than DW against dissipation and fluctuation in the presence of defects or imperfections. To investigate this feature, we inset two equilateral triangular pinning defects (with side length of $0.5W$) in the narrow and wide parts of the nanowire, as shown in Fig. 3. Under the same driving current density, we can observe that the DW is pinned whereas the skyrmion passes successfully the defect region because the pinning effect of the defect on the skyrmion is much weaker than DW. The critical depinning current density of skrymion is 2-4 orders lower than that of DW (see the inset of Fig. 4). Therefore, skyrmion allows a more robust data transmission compared with DW motion.

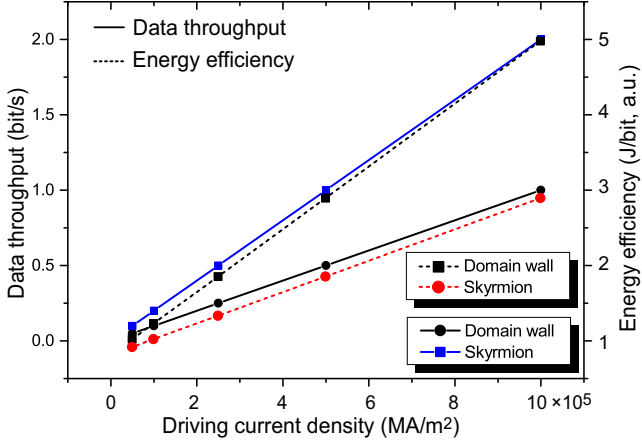

Fig. 5. Comparison of the data throughput and energy efficiency between the skyrmion- and DW-based interconnects.

Furthermore, the size and spacing of skyrmion can be down to several nanometers (~25 nm in our simulations), whereas the spacing between pairs of DWs can hardly be reduced below 30-40 nm (~50 nm in our simulations) [9]. Although the ratio between the velocity and driving current density of skyrmion is approximately as the same order as that of DW when the driving current density is sufficiently large (see Fig .4), smaller size and spacing of skyrmion allow denser data flows than that of DW with the same driving current density, resulting in more energy-efficient or higher throughput data transmission. In our simulations, the skyrmion-based interconnect achieves almost twice larger data throughput (or energy efficiency) than that of the DW-based one, as shown in Fig. 5.

## III. SIMULATION

To validate the functionality and evaluate the performance of the proposed method, we performed micromagnetic studies by solving the Landau-Lifshitz-Gilbert equation, which governs the dynamics of the magnetization $\boldsymbol{m}$,

$$\frac{d\boldsymbol{m}}{dt} = -|\gamma|\boldsymbol{m} \times \boldsymbol{H}_{eff} + \alpha \boldsymbol{m} \times \frac{d\boldsymbol{m}}{dt} + u\boldsymbol{m} \times \left(\frac{\partial \boldsymbol{m}}{\partial x} \times \boldsymbol{m}\right) + \beta u \left(\boldsymbol{m} \times \frac{\partial \boldsymbol{m}}{\partial x}\right) \quad (1)$$

where $\gamma$ is the gyromagnetic ratio, $\alpha$ is the Gilbert damping, $u = \gamma \hbar j P / 2eM_s$, $\hbar$ is the reduced Planck constant, $j$ is the driving current density flowing in the $-x$ direction, $P$ is the spin polarization, $e$ is the electron charge, $M_s$ is the saturation magnetization, and $\beta$ is the non-adiabatic factor [15]. $\boldsymbol{H}_{eff}$ is the effective field including the contributions of Heisenberg exchange, DMI, PMA and demagnetization field.

The nanowire in our simulations is a 1-nm-thick Cobalt on a heavy metal (Pt) substrate. The width (length) of the narrow and wide parts of the nanowire are 25 nm (400 nm) and 50 nm (800 nm). We adopted the following material parameters in our default simulations [10-11]: exchange stiffness $A = 15$ pJm$^{-1}$, damping $\alpha = 0.3$, saturation magnetization $M_s = 580$ kAm$^{-1}$, DMI constant $D = 3$ mJm$^{-3}$, PMA constant $K_u = 0.7$ MJm$^{-3}$, and the spin polarization $P = 1.0$. The Skyrmion number is defined as [10-11],

$$N_{sk} = -\frac{1}{4\pi} \iint dxdy \, \boldsymbol{m} \cdot \left(\frac{\partial \boldsymbol{m}}{\partial x} \times \frac{\partial \boldsymbol{m}}{\partial y}\right) \quad (2)$$

When a DW is converted into a skyrmion, the skyrmion number changes from 0 to 1 and vice versa. This illustrates intuitively a success mutual conservation between a non-topological DW and a topological skyrmion.

## IV. CONCLUSION

We proposed a novel on-chip or (and) inter-chip interconnect design for DW-based systems by exploring the intrinsic merits of skyrmions. The direct conversion between a skyrmion and a DW enables the system to be highly compact. Functionality and performance are evaluated by micromagnetic simulations. The proposed solution provides a compact, robust, energy-efficient and high throughput interconnect, allowing the system to utilize the combined advantages of DW and skyrmion.